\documentclass{amsart}

 \newtheorem{thm}{Theorem}

 \newtheorem{prop}[thm]{Proposition}
 \newtheorem{theorem}[thm]{Theorem}
 \theoremstyle{definition}
 
 \theoremstyle{remark}

 \newcommand{\Real}{\mathbb{R}}
  \newcommand{\Integer}{\mathbb{Z}}

\usepackage{graphicx}

\begin{document}

\title{Existence of travelling waves in discrete sine-Gordon
rings}

\author{Guy Katriel}
\address{Einstein Institute of Mathematics, The Hebrew University of Jerusalem, Jerusalem, 91904, Israel}
\email{haggaik@wowmail.com}

\thanks{Abbreviated title: Travelling waves in sine-Gordon rings}


\dedicatory{}
\thanks{Partially supported by the Edmund Landau
Center for Research in Mathematical Analysis and Related Areas,
sponsored by the Minerva Foundation (Germany).}

\begin{abstract}
We prove existence results for travelling waves in discrete,
damped, dc-driven sine-Gordon equations with periodic boundary
conditions. Methods of nonlinear functional analysis are employed.
Some unresolved questions are raised.
\end{abstract}

\maketitle

\section{Introduction}
The damped, dc-driven discrete sine-Gordon equation, known also as
the driven Frenkel-Kontorova model, with periodic boundary
conditions, arises as a model of many physical systems, including
circular arrays of Josephson junctions, the motions of
dislocations in a crystal, the adsorbate layer on the surface of a
crystal, ionic conductors, glassy materials, charge-density wave
transport, sliding friction, as well as the mechanical
interpretation as a model for a ring of pendula coupled by
torsional springs (we refer to \cite{strunz,watanabe,zheng} and
references therein). This model has thus become a fundamental one
for nonlinear physics, and has been the subject of many
theoretical, numerical and experimental studies. The system of
equations is
\begin{equation}
\label{sg} \phi_j''+\Gamma
\phi_j'+\sin(\phi_j)=F+K[\phi_{j+1}-2\phi_j+\phi_{j-1}],
\;\;\;\forall j\in \Integer
\end{equation}
with the parameters $\Gamma>0, K>0, F>0$, with the periodic
boundary-condition
\begin{equation}\label{bc}
\phi_{j+n}(t)=\phi_j(t)+2\pi m \;\;\;\forall j\in \Integer
\end{equation}
where $m\geq 1$ (we note that in view of the boundary conditions
we are really dealing with an $n$-dimensional system of ODE's
rather than an infinite-dimensional one). In numerical
simulations, as well as in experimental work on systems modelled
by (\ref{sg}),(\ref{bc}), it is observed that solutions often
converge to a travelling wave: a solution satisfying
\begin{equation}\label{pf}
\phi_j(t)=f\Big(t+j\frac{m}{n}T\Big),
\end{equation}
where the waveform $f:\Real\rightarrow\Real$ is a function
satisfying
\begin{equation}\label{pp}
f(t+T)=f(t)+2\pi \;\;\; \forall t\in \Real.
\end{equation}
The velocity of the travelling wave is given by
\begin{equation}
\label{vel}v=\frac{2\pi}{T}.
\end{equation}
However, as has been pointed out in \cite{watanabe}, even the
{\it{existence}} of such a solution, has not been proven, except
for the case of small $K$ in which existence of a travelling wave
for some values of $F$ had been proven in \cite{levi}.

In the `super-damped' case, in which the second-derivative term in
(\ref{sg}) is removed, there are very satisfactory results about
existence and also global stability of travelling waves
(\cite{baesens}, theorem 2). Such results rely strongly on
monotonicity arguments. Recently Baesens and Mackay \cite{mackay}
have managed to extend these arguments to the `overdamped' case of
(\ref{sg}): their results apply when
\begin{equation}\label{bm}
\Gamma>2\sqrt{2K+1},
\end{equation}
and say that there exists a travelling-wave solution which is
globally stable if and only if (\ref{sg}),(\ref{bc}) does not have
stationary solutions. We do not know whether in general the
non-existence of stationary solutions implies the existence of a
travelling wave.

We note that a function $f$ is a waveform if and only if it
satisfies (\ref{pp}) and
\begin{equation}\label{wc}
f''(t)+\Gamma f'(t)
+\sin(f(t))=F+K\Big[f\Big(t+\frac{m}{n}T\Big)-2f(t)+f\Big(t-\frac{m}{n}T\Big)\Big].
\end{equation}

Here we obtain several existence results for travelling-waves
under conditions not covered by the existing work, described
above.

\begin{theorem}
\label{E0}
Fixing any $\Gamma>0$, $K>0$, and
given any velocity $v>0$, there exists a
travelling-wave solution of (\ref{sg}), (\ref{bc})
with velocity $v$ for an appropriate $F>0$.
\end{theorem}

\begin{theorem}
\label{E1} For any $F>1$ there exists a travelling-wave solution
of (\ref{sg}), (\ref{bc}).
\end{theorem}

\begin{theorem}
\label{E2} Assume that $n$ does not divide $m$. Fixing any
$\tilde{F}>0$ and $\tilde{\Gamma}>0$, for all $K$ sufficiently large there
exists a travelling-wave solution of (\ref{sg}), (\ref{bc}) for
any $F\geq\tilde{F}$, $\Gamma\geq \tilde{\Gamma}$.
\end{theorem}

\begin{theorem}
\label{E3} Fixing any $\tilde{F}>0$ and $\tilde{K}>0$, for all
$\Gamma>0$ sufficiently small there exists a travelling-wave
solution of (\ref{sg}), (\ref{bc}) for any $F\geq\tilde{F}$,
$0<K\leq \tilde{K}$.
\end{theorem}

We remark that the assumption that $n$ does not divide $m$ cannot
be removed from theorem \ref{E2}, since if $n$ divides $m$ the
coupling term vanishes and (\ref{wc}),(\ref{pp}) reduce to the
equation of a running solution of a dc-forced pendulum, which,
fixing $\Gamma>0$, is known to have a solution only when $F$
exceeds a positive critical value \cite{miranker}.

It is interesting to note that theorem \ref{E3} demonstrates that
for some parameter ranges there is coexistence of stationary
solutions and travelling waves of (\ref{sg}), (\ref{bc}). Indeed,
it is well known \cite{floria} that, fixing $K$, for $F$
sufficiently small there exist stationary solutions of
(\ref{sg}), (\ref{bc}), and these obviously do not depend on
$\Gamma$. Hence we can take $\Gamma>0$ sufficiently small so that
theorem \ref{E3} ensures also the existence of travelling waves.
This phenomenon cannot happen in the super-damped case, nor in the
overdamped case in which (\ref{bm}) holds, since in
these cases existence of stationary solutions implies that the
$\omega$-limit set of every orbit is contained in the set of
stationary solutions \cite{baesens,mackay}.

Along the way we will prove that
\begin{prop}
\label{bounds} An upper bound for the velocity $v$ of any travelling
wave is given by
\begin{equation}
\label{lub} v<\frac{F}{\Gamma}.
\end{equation}
and a lower bound, in the case $F>1$, is given by
\begin{equation}
\label{lub1}  v > \frac{F-1}{\Gamma}.
\end{equation}
\end{prop}

In the next section we prove the results stated above. In section
\ref{discussion} we discuss the meaning of our results in
connection with existing numerical studies of the discrete
sine-Gordon equation, and point out some further mathematical
questions which arise from our results and remain open.

\section{proofs of the results}

Our method of proof involves re-formulating the problem as a fixed
point problem in a Banach space, and applying results of nonlinear
functional analysis. Our approach is thus close in spirit to
\cite{mirollo}, which deals with travelling waves in globally
coupled Josephson junctions.

We transform the problem (\ref{pp}),(\ref{wc}) by setting
$$f(t)=u(vt)+vt,$$
where the wave-velocity $v$ is defined by (\ref{vel}) and $u$
satisfies:
\begin{equation}\label{per}
u(z+2\pi)=u(z) \;\;\;\forall z\in \Real.
\end{equation}
(\ref{wc}) can then be written as
\begin{eqnarray}\label{wct}
v^2 u''(z)&+&\Gamma v u'(z) +\sin(z+u(z))\nonumber\\&=&F-\Gamma v
+K\Big[u\Big(z+2\pi\frac{m}{n}\Big)-2u(z)+u\Big(z-2\pi\frac{m}{n}\Big)\Big]
\end{eqnarray}
Dividing by $v^2$ and setting
$$\lambda=\frac{1}{v}$$
we re-write (\ref{wct}) in the form
\begin{eqnarray}\label{wf}
u''(z)&+&\lambda \Gamma
u'(z)+\lambda^2\sin(z+u(z))\nonumber\\&=&\lambda^2 F-\lambda
\Gamma +\lambda^2
K\Big[u\Big(z+2\pi\frac{m}{n}\Big)-2u(z)+u\Big(z-2\pi\frac{m}{n}\Big)\Big].
\end{eqnarray}
We note that if $u(z)$ satisfies (\ref{per}),(\ref{wf}) then so
does $\tilde{u}(z)=u(z+c)+c$, for any $c\in \Real$. Thus by
adjusting $c$ we may assume that $u$ satisfies
\begin{equation}\label{zm}
\int_0^{2\pi}{u(s)ds}=0.
\end{equation}
We note now that if $u$ satisfies (\ref{per}),(\ref{wf}), then by
integrating both sides of (\ref{wf}) over $[0,2\pi]$ we obtain
\begin{equation}\label{inte}
F=\frac{\Gamma}{\lambda}+\frac{1}{2\pi}\int_0^{2\pi}{\sin(s+u(s))ds}.
\end{equation}
We can thus re-write (\ref{wf}) as
\begin{eqnarray}\label{wff}
u''(z)+\lambda \Gamma
u'(z)&+&\lambda^2\sin(z+u(z))=\lambda^2\frac{1}{2\pi}\int_0^{2\pi}{\sin(s+u(s))ds}\\&+&\lambda^2
K\Big[u\Big(z+2\pi\frac{m}{n}\Big)-2u(z)+u\Big(z-2\pi\frac{m}{n}\Big)\Big].\nonumber
\end{eqnarray}
Conversely, if $u$ satisfies (\ref{inte}) and (\ref{wff}) then it
satisfies (\ref{wf}). We have thus reformulated our problem as:
find solutions $(\lambda,u)$ of
(\ref{per}),(\ref{zm}),(\ref{inte}),(\ref{wff}). The idea now is
to consider $\lambda$ as a {\it{parameter}} in (\ref{wff}) and try
to find solutions $u$ satisfying
(\ref{per}),(\ref{zm}),(\ref{wff}) and then substitute $\lambda$
and $u$ into (\ref{inte}) to obtain the corresponding value of
$F$.  This is the same idea as used in the numerical method
presented in \cite{strunz}, but here it is used as part of
existence proofs. We claim that
\begin{prop}
\label{exs} For any value $\lambda$, there exists a solution $u$
of (\ref{wff}) satisfying (\ref{per}),(\ref{zm}).
\end{prop}
We note that this proposition immediately implies theorem
\ref{E0}, since given any $v>0$ it shows that we can solve
(\ref{wff}) with $\lambda=\frac{1}{v}$, hence obtain a travelling
wave with velocity $v$, for the value of $F$ given by
(\ref{inte}).

To prove proposition \ref{exs} we will use the Schauder fixed-point theorem. We
denote by $X,Y$ the Banach spaces of real-valued functions
$$X=\{ u\in H^2[0,2\pi] \; | \; u(0)=u(2\pi),\;\;
u'(0)=u'(2\pi),\;\; \int_0^{2\pi}{u(s)ds}=0\},$$
$$Y=\{ u\in L^2[0,2\pi] \;|\; \int_0^{2\pi}{u(s)ds}=0\},$$
with the norm
$$\|u \|_Y=\Big( \frac{1}{2\pi}\int_0^{2\pi}{(u(s))^2ds} \Big)^{\frac{1}{2}},$$
 and by $L_{\lambda}:X\rightarrow Y$ the linear
mapping
$$L_{\lambda}(u)(z)=u''(z)+\lambda \Gamma u'(z)-\lambda^2 K\Big[u\Big(z+2\pi\frac{m}{n}\Big)-2u(z)+u\Big(z-2\pi\frac{m}{n}\Big)\Big].$$
We want to show that this mapping is invertible and derive an
upper bound for the norm of its inverse. Noting that any $u\in X$
can be decomposed in a Fourier series $u(z)=\sum_{l\neq 0} {a_l
e^{ilz}}$ (with $a_{-l}=\overline{a_l}$), we apply $L_{\lambda}$
to the Fourier elements, obtaining,
$$L_{\lambda}(e^{ilz})=\mu_l e^{ilz},$$
where
$$\mu_l=-l^2-2K\lambda^2\Big(\cos\Big(\frac{2\pi ml}{n}\Big)-1\Big)+\lambda
l \Gamma i,$$ so that
\begin{equation}\label{eigb}
|\mu_l|=\Big[\Big(l^2+2K\lambda^2\Big(\cos\Big(\frac{2\pi
ml}{n}\Big)-1\Big)\Big)^2+\lambda^2
l^2\Gamma^2\Big]^{\frac{1}{2}},
\end{equation}
which does not vanish if $\Gamma>0$. Thus the mapping
$L_{\lambda}$ has an inverse satisfying
$L_{\lambda}^{-1}(e^{ilz})=\frac{1}{\mu_l}e^{ilz}$. Since
$L_{\lambda}^{-1}$ takes $Y$ onto $X$, and since $X$ is compactly
embedded in $Y$, we may consider $L_{\lambda}^{-1}$ as a mapping
from $Y$ to itself, in which case it is a {\it{compact}} mapping.
We also note using (\ref{eigb}) that
\begin{equation}\label{star}\|
L_{\lambda}^{-1}\|_{Y,Y}\leq \max_{l\geq 1}{\frac{1}{|\mu_l|}}\leq
\frac{1}{\lambda \Gamma}. \end{equation} We also define the
nonlinear operator $N:Y\rightarrow Y$ by
$$N(u)(z)=-\sin(z+u(z))+\frac{1}{2\pi}\int_0^{2\pi}{\sin(s+u(s))ds}.$$
It is easy to see that $N$ is continuous, and that the range of
$N$ is contained in a bounded ball in $Y$, indeed we have
$$\| \sin(z+u(z)) \|_{L_2}=\Big( \frac{1}{2\pi}\int_0^{2\pi}{(\sin(s+u(s)))^2ds}\Big)^{\frac{1}{2}}\leq 1,$$
and since $N(u)$ is the orthogonal projection of $-\sin(z+u(z))$
into $Y$, we have
\begin{equation}\label{bnd}
\| N(u) \|_Y \leq 1 \;\;\; \forall u\in Y.
\end{equation}

We can now rewrite the problem (\ref{per}),(\ref{zm}),(\ref{wff}) as the
fixed-point problem:
\begin{equation}\label{fixp}
u=\lambda^2 L_{\lambda}^{-1}\circ N(u).
\end{equation}
The operator on the right-hand side is compact by the compactness
of $L_{\lambda}^{-1}$, and has a bounded range by
(\ref{star}),(\ref{bnd}), so that Schauder's fixed-point theorem
implies that (\ref{fixp}) has a solution, proving proposition
\ref{exs} (we note that by a simple bootstrap argument a solution
in $Y$ is in fact smooth). Moreover, defining
$$\Sigma =\{ (\lambda,u)\in
[0,\infty)\times Y\;|\; u=\lambda^2 L_{\lambda}^{-1}\circ
N(u)\},$$ Rabinowitz's continuation theorem \cite{rabinowitz}
implies that the connected component of $\Sigma$ containing
$(\lambda,u)=(0,0)$, which we denote by $C$, is unbounded in
$[0,\infty)\times Y$. Since for any $\lambda_0>0$ we have, from
(\ref{bnd}), (\ref{fixp}), the bound $\| u\|_Y\leq
\frac{\lambda_0}{\Gamma}$ for solutions $(\lambda,u)$ of
$(\ref{fixp})$ with $\lambda\in[0,\lambda_0]$, the unboundedness
of the set $C$ must be in the $\lambda$-direction, that is, there
exist $(\lambda,u)\in C$ with arbitrarily large values of
$\lambda$.

We can now consider the right-hand side of
(\ref{inte}) as a functional on $[0,\infty)\times Y$:
\begin{equation}
\label{dphi}\Phi(\lambda,u)=\frac{\Gamma}{\lambda}+\frac{1}{2\pi}\int_0^{2\pi}{\sin(s+u(s))ds},
\end{equation}
and our strategy in proving theorems \ref{E1}-\ref{E3} is to prove
solvability of the equation
\begin{equation}
\label{cent}\Phi(\lambda,u)=F,\;\;\;(\lambda,u)\in \Sigma
\end{equation}
(in fact we shall prove solvability of (\ref{cent}) with $\Sigma$
replaced by $C\subset \Sigma$). We note that by the boundedness of
the sine function we have
\begin{equation}\label{lim0}
\lim_{\lambda\rightarrow 0+,\; (\lambda,u)\in
C}{\Phi(\lambda,u)}=+\infty,
\end{equation}
\begin{equation}\label{limi}
\limsup_{\lambda\rightarrow +\infty,\; (\lambda,u)\in
C}{\Phi(\lambda,u)}\leq 1.
\end{equation}
Since $C$ is a connected set and $\Phi$ is continuous,
(\ref{lim0}) implies that
\begin{prop} \label{interval}
For any $F$ satisfying
\begin{equation}
\label{dof} F>\underline{F}\equiv\inf_{(\lambda,u)\in
C}{\Phi(\lambda,u)},
\end{equation}
there exists a travelling wave.
\end{prop}
Since (\ref{limi}) implies that $\underline{F}\leq 1$, this proves
theorem \ref{E1}.

\vspace{0.4cm}

We now prove the lower and upper bounds for the velocities of
travelling waves given in proposition \ref{bounds}. These follow
from (\ref{vel}),(\ref{inte}) and

\begin{prop}
\label{bounds1} For any $(\lambda,u)\in \Sigma$ with $\lambda>0$
we have
$$0<\frac{\Gamma}{\lambda}<\Phi(\lambda,u)<\frac{\Gamma}{\lambda}+1.$$
\end{prop}
The upper bound follows immediately from the definition
(\ref{dphi}) of $\Phi(\lambda,u)$ since
$\frac{1}{2\pi}\int_0^{2\pi}{\sin(s+u(s))ds}<1$. The lower bound
follows from the claim that
\begin{equation} \label{poc}
(\lambda,u)\in \Sigma\;
\Rightarrow\;\int_0^{2\pi}{\sin(s+u(s))ds}>0.
\end{equation}
To prove this claim we multiply (\ref{wff}) by $1+u'(z)$ and
integrate over $[0,2\pi]$, noting that
\begin{eqnarray*}
\int_0^{2\pi}{u\Big(s+2\pi\frac{m}{n}\Big)u'(s)ds}&=&\int_0^{2\pi}{u(s)
u'\Big(s-2\pi\frac{m}{n}\Big)ds}\nonumber\\&=&-\int_0^{2\pi}{u'(s)
u\Big(s-2\pi\frac{m}{n}\Big)ds},
\end{eqnarray*} so that we obtain
$$(\lambda,u)\in\Sigma\Rightarrow \Gamma\frac{1}{2\pi}\int_0^{2\pi}{(u'(s))^2ds}=\lambda
\frac{1}{2\pi} \int_0^{2\pi}{\sin(s+u(s))ds}.$$ This proves
(\ref{poc}) since the left-hand side is non-negative and cannot
vanish unless $u\equiv 0$, but $(\lambda,0)\not\in\Sigma$ for
$\lambda>0$.

 \vspace{0.4cm}

We now turn to the proof of theorem \ref{E2}.
\begin{prop}
\label{kl} Assume $n$ does not divide $m$ and $\tilde{\Gamma}>0$.
Given any  $\lambda_0>0$ and $\epsilon>0$, there exists $K_0$ such
that for $K\geq K_0$, $\Gamma\geq \tilde{\Gamma}$ we have that
\begin{equation}\label{ikl}
\Big| \frac{1}{2\pi}\int_0^{2\pi}{\sin(s+u(s))ds} \Big|
<\epsilon\;\;\;if\; (\lambda_0,u)\in \Sigma.
\end{equation}
\end{prop}
To see that proposition \ref{kl} implies theorem \ref{E2}, we fix
some $\tilde{F}>0$, $\tilde{\Gamma}>0$, and assume $\Gamma\geq
\tilde{\Gamma}$. We choose $\lambda_0>\frac{\Gamma}{\tilde{F}}$
and set $\epsilon=\tilde{F}-\frac{\Gamma}{\lambda_0}$. We then
choose $K_0$ according to proposition \ref{kl}, so that
(\ref{ikl}) holds, which implies that when $K\geq K_0$ we have
$\Phi(\lambda_0,u)<\tilde{F}$ for any $u$ with $(\lambda_0,u)\in
C$. Thus $\underline{F}<\tilde{F}$, where $\underline{F}$ is
defined by (\ref{dof}), so proposition \ref{interval} implies the
existence of a travelling wave for any $F\geq \tilde{F}$.

 \vspace{0.4cm} We now prove proposition \ref{kl}. Let
$\lambda_0>0$ and $\epsilon>0$ be given. Assume $(\lambda_0,u)\in
\Sigma$, so that (\ref{fixp}) holds with $\lambda=\lambda_0$. Let
$(m,n)$ denote the greatest common divisor of $m,n$ and let
$$p=\frac{m}{(m,n)},\;\; q=\frac{n}{(m,n)}.$$
Since we assume $n$ does not divide $m$ we have $q\geq 2$.
 Let $Y_0$ be the subspace of $Y$ consisting of
$\frac{2\pi}{q}$-periodic functions, and let $Y_1$ be its
orthogonal complement in $Y$. We denote by $P$ the orthogonal
projection of $Y$ to $Y_0$. Setting $$u_0=P(u),
\;\;u_1=(I-P)(u),$$ we have $u=u_0+u_1$ with $u_0\in Y_0$,
$u_1\in Y_1$. Applying $P$ and $I-P$ to (\ref{fixp}), and noting
that $L_{\lambda}$ commutes with $P$, we have
\begin{equation}\label{pro1}
u_0=\lambda_0^2 L_{\lambda_0}^{-1}\circ P\circ N(u_0+u_1)
\end{equation}
\begin{equation}\label{pro2}
u_1=\lambda_0^2 L_{\lambda_0}^{-1}\circ (I-P) \circ N(u_0+u_1)
\end{equation}
We will now use (\ref{eigb}) to derive a bound for $\|
L_{\lambda_0}^{-1}|_{Y_1}\|_{Y_1,Y_1}$ which goes to $0$ as
$K\rightarrow \infty$. We note that
\begin{equation}
\label{sstar}\| L_{\lambda_0}^{-1}|_{Y_1}\|_{Y_1,Y_1}\leq
\max_{l\geq 1,\; q \not\;|\; l}{\frac{1}{|\mu_l|}}, \end{equation}
so we need to find lower bounds for the $|\mu_l|$'s for which $q$
does not divide $l$. We define
$$\rho=\max_{l\geq 1,\; q \not\;|\; l} {\cos\Big( \frac{2\pi
pl}{q}\Big)}$$ and note that since $p,q$ are coprime we have
$\rho<1$.

 We define
$$\alpha= 2K\lambda_0^2( 1-\rho)
-\sqrt{K},$$ and we shall henceforth assume that $K$ is
sufficiently large so that $\alpha>0$. For each $l\geq 1$ we have
either $l^2<\alpha$ or $l^2\geq \alpha$, and we treat each of
these cases separately.

\noindent (1) In case $l^2<\alpha$, we have
$$l^2+2K\lambda_0^2(\rho -1)< -\sqrt{K},$$ and
by the definition of $\rho$
$$\cos \Big(\frac{2\pi ml}{n}\Big)=\cos \Big(\frac{2\pi pl}{q}\Big) \leq \rho,$$
so that
$$l^2+2K\lambda_0^2\Big(\cos\Big(\frac{2\pi
ml}{n}\Big)-1\Big)<-\sqrt{K},$$ which by (\ref{eigb}) implies
\begin{equation}\label{cc1}
|\mu_l|> \sqrt{K}.
\end{equation}

\noindent (2)  In case $l^2\geq \alpha$, we have, since
(\ref{eigb}) implies $|\mu_l|>\lambda_0\Gamma l$:
\begin{equation}\label{cc2}
|\mu_l|\geq \lambda_0 \Gamma \sqrt{\alpha}\geq
\lambda_0\tilde{\Gamma}\sqrt{\alpha}= \lambda_0\tilde{\Gamma}
\Big[2K\lambda_0^2( 1-\rho) -\sqrt{K}\Big]^{\frac{1}{2}}.
\end{equation}
From (\ref{cc1}),(\ref{cc2}) we obtain that
$\lim_{K\rightarrow\infty}{|\mu_l|}=+\infty$ uniformly with
respect to $l\geq 1$ which are not multiples of $q$, hence by
(\ref{sstar})
$$\lim_{K\rightarrow \infty }\|
L_{\lambda_0}^{-1}|_{Y_1}\|_{Y_1,Y_1}=0.$$
In particular we may
choose $K_0$ such that for $K\geq K_0$ we will have
$$\| L_{\lambda_0}^{-1}|_{Y_1}\|_{Y_1,Y_1}<\frac{\epsilon}{\lambda_0^2}.$$
By (\ref{pro2}) and (\ref{bnd}) this implies
 \begin{equation}
\label{uap} \| u_1 \|_Y \leq \epsilon.
\end{equation}
Thus \begin{eqnarray}\label{bin} \Big|
\frac{1}{2\pi}\int_0^{2\pi}{\sin(s+u(s))ds} \Big| &\leq& \Big|
\frac{1}{2\pi}\int_0^{2\pi}{\sin(s+u_0(s))ds} \Big| \nonumber
\\&+& \Big| \frac{1}{2\pi}\int_0^{2\pi}{[\sin(s+
u(s))-\sin(s+u_0(s))]ds}\Big|\nonumber\\ &\leq& \Big|
\frac{1}{2\pi}\int_0^{2\pi}{\sin(s+u_0(s))ds} \Big| +
\frac{1}{2\pi}\int_0^{2\pi}{|u_1(s)|ds}
\end{eqnarray}
From (\ref{uap}) and the Cauchy-Schwartz inequality we have
\begin{equation}
\label{nt}\frac{1}{2\pi}\int_0^{2\pi}{|u_1(s)|ds}\leq
\frac{1}{\sqrt{2\pi}}\Big( \int_0^{2\pi}{(u_1(s))^2
ds}\Big)^{\frac{1}{2}}\leq \epsilon. \end{equation} From
trigonometry we have
$$\int_0^{2\pi}{\sin(s+u_0(s))ds} =
\int_0^{2\pi}{\sin(s)\cos(u_0(s))ds}+\int_0^{2\pi}{\cos(s)\sin(u_0(s))ds},$$
but the functions $\cos(u_0(s))$, $\sin(u_0(s))$ are
$\frac{2\pi}{q}$-periodic with $q\geq 2$, which implies that they
are orthogonal to $\cos(s)$, $\sin(s)$, so that we have
$$\int_0^{2\pi}{\sin(s+u_0(s))ds}=0,$$
which together with (\ref{bin}) and (\ref{nt}) implies
(\ref{ikl}), concluding the proof of proposition \ref{kl}.

 \vspace{0.4cm}
We now turn to the proof of theorem \ref{E3}. We first note that
from (\ref{eigb}) we have
$$|\mu_l|\geq \Big| l^2+2K\lambda^2\Big(\cos\Big(\frac{2\pi m
l}{n}\Big) -1\Big)\Big|,$$ so that if we assume
$$0< \lambda <\lambda_0 = \frac{1}{\sqrt{8\tilde{K}}}\leq\frac{1}{\sqrt{8K}}$$
then we have $|\mu_l|>\frac{1}{2}$ for all $l\geq 1$, hence $\|
L_{\lambda}^{-1}\|_{Y,Y}< 2$, independently of $\Gamma$. From
(\ref{fixp}) we thus have
$$(\lambda,u)\in \Sigma,\;0<\lambda<\lambda_0 \; \Rightarrow\; \| u
\|_Y < 2\lambda^2.$$ We now choose $\lambda_1\leq \lambda_0$ so
that $2\lambda_1^2\leq \frac{1}{2}\tilde{F}$. Thus
$(\lambda_1,u)\in \Sigma$ implies that
\begin{eqnarray}
\label{ccc} &&\Big|
\frac{1}{2\pi}\int_0^{2\pi}{\sin(s+u(s))ds}\Big| =
\Big|\frac{1}{2\pi}\int_0^{2\pi}{[\sin(s+u(s))-\sin(s)]ds}\Big|\\&\leq&
 \frac{1}{2\pi}\int_0^{2\pi}{|u(s)|ds}\leq \| u \|_Y<
 2\lambda_1^2\leq \frac{1}{2}\tilde{F}.\nonumber
 \end{eqnarray}
Finally, we choose $\Gamma_0$ so that
\begin{equation}\label{ddd}
\frac{\Gamma_0}{\lambda_1}<\frac{1}{2}\tilde{F}.
\end{equation}
(\ref{ccc}), (\ref{ddd}) thus imply that when $0<\Gamma<\Gamma_0$
$$(\lambda_1,u)\in \Sigma \; \Rightarrow\; \Phi(\lambda_1,u)<\tilde{F},$$
so that we have $\underline{F}<\tilde{F}$, where $\underline{F}$
is defined by (\ref{dof}), hence proposition \ref{interval}
implies the existence of a travelling wave for any $F\geq
\tilde{F}$.

\section{Discussion and further questions}
\label{discussion}

In the numerical and experimental explorations of the dynamics of
sine-Gordon rings \cite{strunz,watanabe,zheng}, a useful method of
representation consists in displaying the velocity-force
characteristic. In the case of the travelling waves studied here,
since the velocity of the waves is given by $v=\frac{1}{\lambda}$,
the velocity-force characteristic is the subset of the
$(F,v)$-plane given by
$$\Big{\{} \Big(\Phi\Big(\frac{1}{v},u\Big),v\Big)\Big)
\;\Big|\; \Big(\frac{1}{v},u\Big)\in\Sigma \Big{\}},$$ where the
set $\Sigma$ and the functional $\Phi$ are as defined in the
previous section. Examining the velocity-force characteristic as
numerically computed in \cite{strunz} (fig.2), we see that $F$ is
a non-monotone function of $v$. This means that for some values of
$F$ equation (\ref{cent}) has more than one solution, or in other
words that there exist multiple travelling waves with different
velocities for the same value of $F$. On the other hand, the fact
that according to the available numerical evidence $F$ is a
function of $v$ leads us to conjecture that, fixing
$\Gamma>0,K>0$, for each given velocity $v>0$ there is a unique
travelling wave with velocity $v$, for an appropriate $F$. Thus,
our conjecture is that uniqueness holds in theorem \ref{E0}, or in
other words that the fixed point problem (\ref{fixp}) always has a
unique solution. Let us note that for $0<\lambda<\Gamma$ it is
easy to show, using (\ref{star}), that the right-hand side of
(\ref{fixp}) is a contraction from $Y$ to itself, hence we may
replace the use of the Schauder fixed-point theorem by the Banach
contraction-mapping principle, which implies uniqueness. Thus, at
least for velocities $v>\frac{1}{\Gamma}$, we have uniqueness in
theorem \ref{E0}. However for lower velocities this argument does
not work so a proof of the above conjecture will require some new
idea.

The non-uniqueness of travelling waves for some values of the
parameters $\Gamma,K,F$, mentioned above, implies important
consequences for the dynamics of the discrete sine-Gordon ring,
such as instability of some of the travelling waves, and
bistability of travelling waves leading to hysteresis as the force
$F$ is varied. It would be interesting to determine whether these
phenomena can occur in the large $K$ and the small $\Gamma$
regimes for which existence of a travelling wave has been proved
in theorems \ref{E2},\ref{E3}. Moreover, stationary solutions and
travelling waves do not exhaust the dynamical repertoire of the
discrete sine-Gordon equations in the under-damped case:
quasi-periodic and chaotic behavior is reported
\cite{strunz,strunz1,zheng}. An interesting question is to
determine conditions on the parameters which ensure that there
exists at least one {\it{locally asymptotically stable}}
travelling wave.

Returning to the issue of existence of travelling waves, which has
been the focus of our investigation, we note an intriguing
question which arises from our results, and remains unanswered.

Let us define, for fixed $\Gamma>0$, $K>0$
$$F_0(\Gamma,K)= \inf \{ F\geq 0\; |\; {\mbox{a travelling wave of (\ref{sg}), (\ref{bc})
exists}} \}.$$ Theorem \ref{E1} implies that $F_0(\Gamma,K)\leq 1$
for all $\Gamma>0,K>0$. Theorem \ref{E2} implies that, when $n$
does not divide $m$, $\lim_{K\rightarrow
\infty}{F_0(\Gamma,K)}=0$. Theorem \ref{E3} implies that
$\lim_{\Gamma\rightarrow 0}{F_0(\Gamma,K)}=0$. Is it true, though,
that for each $\Gamma>0,K>0$ we have $F_0(\Gamma,K)>0$? In other
words, is it always true that (fixing $\Gamma>0,K>0)$ for
sufficiently small $F>0$ a travelling wave does not exist? We have
not been able to prove or disprove this conjecture, and can only
offer the following remarks:

\noindent (i) If $\Gamma,K$ satisfy (\ref{bm}), then indeed
$F_0(\Gamma,K)>0$, since for sufficiently small $F$ there exists a
stationary solution of (\ref{sg}),(\ref{bc}), so the results of
\cite{mackay} imply that no travelling wave exists for small $F$.
However, as we have remarked, theorem \ref{E3} shows that in
general travelling waves and stationary solutions may coexist.

\noindent (ii) If $n$ divides $m$ then, as was noted in the
introduction, the existence of travelling waves reduces to that of
running solutions of the forced pendulum, hence it is well known
that $F_0(\Gamma,K)>0$ for all $\Gamma,K$. However, in the case
that $n$ divides $m$ we also have noted also that the result of
theorem \ref{E2} does not hold, so that the case that $n$ divides
$m$ is rather special and may not be indicative of the general
case.

\noindent (iii) The conjecture that $F_0(\Gamma,K)>0$ is supported
by the notion of `pinning' - the phenomenon whereby travelling
waves are unable to propagate in discrete systems when the applied
force is small. However, whether this effect indeed holds in
general in underdamped systems (as opposed to the overdamped case
- see (i) above) is unclear to the best of our knowledge.
Moreover, it is conceivable that $F_0(\Gamma,K)=0$ but pinning
still occurs - if for small $F$ a travelling wave exists but is
unstable.

\noindent (iv) Since, by proposition \ref{bounds1}, we have
$\Phi(\lambda,u)>0$ for all $(\lambda,u)\in \Sigma$, $\lambda>0$,
we have that $F_0(\Gamma,K)=0$ if and only if
$$\liminf_{\lambda\rightarrow+\infty,\;(\lambda,u)\in\Sigma}{\Phi(\lambda,u)}=0.$$
Thus, determining whether the above equality can hold could be a
route to resolving our question, but we have not been able to do
so.

\vspace{0.4cm} We conclude with one more question: clarify the
connection, if any, between the travelling waves obtained in
\cite{levi} for small values of $K>0$ and those obtained by us for
large values of $K$ in theorem \ref{E2}.


\begin{thebibliography}{9}

\bibitem{baesens} C. Baesens \& R.S. MacKay, {\it{Gradient dynamics of tilted Frenkel-Kontorova models}},
Nonlinearity {\bf{11}} (1998), 949-964.

\bibitem{mackay} C. Baesens \& R.S. MacKay, {\it{A novel preserved partial order for cooperative networks of units
with overdamped second-order dynamics and application to tilted
Frenkel Kontorova chains }}, Nonlinearity {\bf{17}} (2004),
567--580.

\bibitem{floria} L.M. Flor\'ia  \& J.J. Mazo, {\it{Dissipative dynamics of the Frenkel-Kontorova model}},
 Adv. Phys. {\bf{45}} (1996), 505-598

\bibitem{levi} M. Levi, {\it{Dynamics of discrete Frenkel-Kontorova
models}}, in {\it{Analysis, et cetera}}, ed. P. Rabinowitz \& E.
Zehnder, Academic Press (Boston), 1990.

\bibitem{miranker} M. Levi, F.C. Hoppensteadt \& W.L Miranker,
{\it{Dynamics of the Josephson junction}}, Quarterly, Appl. Math.
{\bf{36}} (2), 167-198.

\bibitem{mirollo} R. Mirollo, N. Rosen, Existence, {\it{Uniqueness, and Nonuniqueness of Single-Wave-Form Solutions
to Josephson Junction Systems}}, SIAM J. Appl. Math {\bf{60}}
(2000), 1471-1501.

\bibitem{rabinowitz} P. Rabinowitz, {\it{Some global results for
nonlinear eigenvalue problems}}, J. Functional Analysis {\bf{7}}
(1971), 487-513.

\bibitem{strunz} T. Strunz \& F.J.Elmer, {\it{Driven Frenkel-Kontorova model: I. Uniform sliding states and
dynamical domains of different particle densities}}, Phys. Rev. E
{\bf{58}} (1998), 1601-1611.

\bibitem{strunz1} T. Strunz \& F.J.Elmer, {\it{Driven Frenkel-Kontorova model: II. Chaotic sliding and nonequilibrium melting and freezing}}, Phys. Rev. E
{\bf{58}} (1998), 1612-1620.

\bibitem{watanabe} S. Watanabe, H.S.J van der Zant, S. Strogatz, \& T.P. Orlando,
{\it{Dynamics of circular arrays of Josephson junctions and the
discrete sine-Gordon equation}}, Physica D {\bf{97}} (1996),
429-470.

\bibitem{zheng} Z. Zheng, B. Hu \& G. Hu,
{\it{Resonant steps and spatiotemporal dynamics in the damped
dc-driven Frenkel-Kontorova chain}}, Physical Review B {\bf{58}}
(1998), 5453-5461.


\end{thebibliography}
\end{document}